\documentclass[aps,showpacs,amsmath,amssymb,amsfonts]{revtex4}

\usepackage{graphicx}
\usepackage{dcolumn}
\usepackage{bm}
\usepackage{hyperref}

\newcommand{\beq}{\begin{equation}}
\newcommand{\eeq}{\end{equation}}
\newcommand{\beqa}{\begin{eqnarray}}
\newcommand{\eeqa}{\end{eqnarray}}
\newcommand{\beqar}{\begin{eqnarray*}}
\newcommand{\eeqar}{\end{eqnarray*}}
\newcommand{\bra}[1]{\mbox{$\left\langle{#1}\right|$}}
\newcommand{\ket}[1]{\mbox{$\left|{#1}\right\rangle$}}
\newcommand{\diracsp}[2]{\mbox{$\langle{#1}|{#2}\rangle$}}

\newcommand{\g}{\gamma}

\newcommand{\Dif}[2]{\frac{\mathrm{d}^{#2}}{\mathrm{d}#1^{#2}}}

\def\I{{\rm i}}

\def\e{{\rm e}}

\newcounter{saveeqn}

\begin{document}

\title{Quantum Mechanics in the general quantum systems (VI):\\
exact series solution of stationary Schr\"odinger's equation}
\author{An Min Wang}
\email{anmwang@ustc.edu.cn}
\affiliation{%
Quantum Theory Group, Department of Modern Physics, University of
Science and Technology of China, Hefei, 230026, P.R.China
}%
\homepage{http://qtg.ustc.edu.cn}%

\begin{abstract}
We obtain a complete series solution of stationary Schr\"odinger's
equation in the general quantum systems. It is exact in the sense
that any approximation means is not used, or that the whole
corrections or contributions from all order perturbations are
involved if the perturbation concept is introduced. Furthermore, the
useful forms of our exact solution and a new expression the complete
Green operator are given out. As a universal and analytical
solution, it is helpful for the theoretical derivations and
practical calculations in quantum theory.
\end{abstract}


\maketitle

Can we know an exact solution of stationary Schr\"odinger's equation
(SSE) \cite{seq} if an arbitrary time-independent Hamiltonian in a
general quantum system is given? The answer from the known theory of
quantum mechanics is {\bf NO}. But, our answer based on this article
is {\bf YES}.

This problem is both basic in the theory of quantum mechanics and
central in the applications of quantum mechanics. If the exact
solution of SSE is known, then, the quantum state at any time is
able to be exactly obtained for a given initial state, the different
probabilities, the overlaps between the states and the expectation
values of physical quantities are able to be expressed exactly and
calculated precisely, the non-perturbation frameworks of quantum
mechanics is possible to be built, and some important conclusions
and theorems should be able be derived out and proved. However, the
interesting on it may be getting a little weak because that this
problem is too old, too difficult, and on the other hand, new
developments and applications of quantum mechanics are very hot. At
present, the exact solutions of SSE only exists in a few simpler
quantum systems. For a general quantum system, this problem has not
been solved since quantum mechanics emerged. In fact, this unsolved
problem is an important origin of a lot of difficulties in theory
developments and practical applications of quantum mechanics. It is
unnecessary to say how important and how interesting of this problem
because anyone working on quantum physics knows its answer. In this
article, let us to study this problem in order to obtain an exact
series solution of stationary Schr\"odinger's equation in the
general quantum systems.

Barring a few exceptions, the Schr\"odinger equation is always
solved in a particular basis or representation. For a general
quantum system, the form of its Hamiltonian $H$ is not known. Thus,
the chosen representation had better to be spanned by the set of
orthogonal, normalization and complete eigenvectors of its solvable
part, just as done in the perturbation theory and Green function
method. That is, we split the Hamiltonian into two parts: \beq
H=H_0+H_1\eeq where the eigenvalues problem of $H_0$
\beq\label{h0eeq} H_0\ket{\Phi_\g}=E_\g \ket{\Phi_\g} \eeq is
solvable. In other words, the eigenvectors $\ket{\Phi_\g}$ and the
corresponding eigenvalues $E_\g$ of $H_0$ can be found or known, and
the so-called $H_0$ representation is constructed by the set
$\{\ket{\Phi_\g} \}$, where $\gamma$ takes over all possible indexes
of energy levels. However, different from the usual perturbation
theory, we do not introduce the perturbation concept. That is, it is
unnecessary that $H_1$ is small enough compared with $H_0$, or $H_1$
is proportional to a very small perturbed parameter, because it will
be seen that no perturbed parameter appears obviously in our
solution, or all order corrections and contributions coming from
$H_1$ have been included for the case that $H_1$ is indeed small
enough compared with $H_0$. Of course, for this typical case, one
also can take up some order approximation of perturbation in our
following solution. However, the result is not trivially equivalent
to the perturbative solution because that the reasonably dynamical
rearrangement feature of our solution leads to the effects from the
different cut approximation.

It must be emphasized that our method to split the Hamiltonian of
system into two parts has such a principle that $H_0$ matrix is
diagonal and $H_1$ matrix is off-diagonal in the representation
spanned by $\{\ket{\Phi_\g} \}$ , that is \beqa
\bra{\Phi_\g}H_0\ket{\Phi_{\g^{\prime}}}=E_{\g}\delta_{\g\g^{\prime}},\qquad
\bra{\Phi_\g}H_1\ket{\Phi_{\g^{\prime}}}=\left\{\begin{array}{ll}g^{\g\g^{\prime}}&
({\rm if} \quad \g\neq\g^\prime)\\
g^{\g\g}=0 & ({\rm if}\quad \g=\g^\prime)\end{array}\right.\eeqa It
is easy to prove that we always make such a division. If there are
the non-vanishing diagonal elements of $H_1$ matrix in the first
division, we then redivide $H_1$ matrix so that all of its
non-vanishing diagonal elements are subtracted and then added to the
$H_0$ matrix in the first division, while all of its off-diagonal
elements is taken as a new $H_1$ matrix. It is obvious that the
redivision result obeys our above principle. At this time, the
eigenvalues of new $H_0$ are added by the non-vanishing diagonal
elements of the original $H_1$ matrix, but the eigenvectors of new
$H_0$ are invariant.£© Note that we do not consider the case that
all off-diagonal elements of $H_1$ matrix are zero. It is trivial
because that the eigenvalue problem of the Hamiltonian has been
solved in this case.

It should point out that an obvious reason that we split the total
Hamiltonian matrix into diagonal part and off-diagonal part rather
than the nonperturbed part and the perturbed part marked by a
perturbed parameter in the normal perturbation theory is: {\em we
should, even have to, give up the perturbation concept in the
beginning, otherwise we will stick in the difficulty same as the
normal perturbation theory}. Moreover, in our point of view, {\em
introducing the perturbed parameter and expanding according to its
power is obviously a transcendent input, and this input covers the
inherent feature that the perturbation series is multiple and
excludes some important cases that the interaction strength is not
very weak}. In fact, giving up the perturbation concept is the most
important reason why our exact series solution of SSE has new
physics content.

We first directly express our exact series solution of SSE just like
an {\it Ansatz} so that it can be understood conveniently and
easily. Then we prove it, and its derivation is arranged behind the
verification.

In a general quantum system, its stationary Schr\"odinger's equation
reads \cite{diracpqm} \beq\label{HEE}
H\ket{\Psi_\g}=\widetilde{E}_\g\ket{\Psi_\g}\eeq The exact series
solution without any approximation of this equation is \beq
\label{HEVector}\ket{\Psi_\g}=\ket{\Phi_\g}+\sum_{\g^\prime \neq
 \g} Q_{\g^\prime\g}\ket{\Phi_{\g^\prime}}\eeq
where \vskip -0.1in \beq\label{Qdefinition}
Q_{\g^\prime\g}=\frac{g^{\g^\prime\g}}{\widetilde{E}_\g-E_{\g^\prime}}
+\sum_{l=1}^\infty\sum_{\g_1,\g_2,\cdots,\g_l\neq
\g}\frac{g^{\g^\prime\g_1}g^{\g_1\g_2}\cdots
g^{\g_l\g}}{(\widetilde{E}_{\g}-E_{\g^\prime})(\widetilde{E}_{\g}
-E_{\g_1})(\widetilde{E}_{\g}-E_{\g_2})\cdots(\widetilde{E}_{\g}-E_{\g_l})}
\eeq While the corresponding eigenvalue $\widetilde{E}_{\g}$ as well
as the quantity $\widetilde{E}_{\g}$ appearing in $Q_{\g^\prime\g}$
definition is a real solution of following equation
\beq\label{ourHEVE}
R_\g(E_\gamma-\widetilde{E}_\gamma)=\widetilde{E}_\gamma- E_\g\eeq
Here, $R_\g(z)$ is called as a kernel function in our expression
defined by \beq
R_\g(z)=\sum_{l=1}^\infty\sum_{\g_1,\g_2,\cdots,\g_l\neq
\g}\frac{g^{\g\g_1}g^{\g_1\g_2}\cdots
g^{\g_l\g}}{(E_{\g}-E_{\g_1}-z)(E_{\g}-E_{\g_2}-z)\cdots(E_{\g}-E_{\g_l}-z)}
\eeq  Note that the convergence of our series solution is implied by
the existence of real solution of Eq.(\ref{ourHEVE}). Otherwise we
can not have above solution of SSE.

Now let us directly prove that our solution is correct. Firstly,
acting the total Hamiltonian on the right side of
Eq.(\ref{HEVector}), we get \beqa\label{vstep1}
H\left(\ket{\Phi_\g}+\sum_{\g^\prime \neq
 \g}
 Q_{\g^\prime\g}\ket{\Phi_{\g^\prime}}\right)&=&E_\g\ket{\Phi_\g}+\sum_{\g^\prime \neq
 \g}
 Q_{\g^\prime\g}E_{\g^\prime}\ket{\Phi_{\g^\prime}}+\sum_{\g^\prime \neq
 \g}g^{\g^\prime\g}\ket{\Phi_{\g^\prime}}+\sum_{\g^{\prime\prime}}\sum_{\g^\prime \neq
 \g}Q_{\g^\prime\g}g^{\g^{\prime\prime}\g^\prime}\ket{\Phi_{\g^{\prime\prime}}}\eeqa
Secondly, we divide the fourth term into two parts with respect to
$\g^{\prime\prime}=\g$ and $\g^{\prime\prime}\neq\g$ \beqa
\label{secondstep} \sum_{\g^{\prime\prime}}\sum_{\g^\prime \neq
 \g}Q_{\g^\prime\g}g^{\g^{\prime\prime}\g^\prime}\ket{\Phi_{\g^{\prime\prime}}}&=&\sum_{\g^\prime \neq
 \g}g^{\g\g^\prime}Q_{\g^\prime\g}\ket{\Phi_{\g}}+\sum_{\g^\prime,\g^{\prime\prime} \neq
 \g}g^{\g^{\prime\prime}\g^\prime}Q_{\g^\prime\g}\ket{\Phi_{\g^{\prime\prime}}}\eeqa
Its first term can be rewritten as \beqa \sum_{\g^\prime \neq
 \g}g^{\g\g^\prime}Q_{\g^\prime\g} &=&\sum_{\g^\prime\neq \g}
 \frac{g^{\g\g^\prime}g^{\g^\prime\g}}{\widetilde{E}_\g-E_{\g^\prime}}
 +\sum_{l=1}^\infty\sum_{\g^\prime,\g_1,\g_2,\cdots,\g_l\neq
\g}\frac{g^{\g\g^\prime}g^{\g^\prime\g_1}g^{\g_1\g_2}\cdots
g^{\g_l\g}}{(\widetilde{E}_{\g}-E_{\g^\prime})(\widetilde{E}_{\g}-E_{\g_1})
(\widetilde{E}_{\g}-E_{\g_2})\cdots(\widetilde{E}_{\g}-E_{\g_l})}
\eeqa By redenoting $\g^\prime$ by $\g_1$ in the first summation,
and $\{\g^\prime,\g_1,\cdots,\g_l\}$ by
$\{\g_1,\g_2,\cdots,\g_{l+1}\}$ in the second summation, the above
equation becomes \beqa \sum_{\g^\prime \neq
 \g}g^{\g\g^\prime}Q_{\g^\prime\g} &=&\sum_{\g_1\neq \g}\frac{g^{\g\g_1}g^{\g_1\g}}{\widetilde{E}_\g-E_{\g_1}}
 +\sum_{l=1}^\infty\sum_{\g_1,\g_2,\cdots,\g_{l+1}\neq
\g}\frac{g^{\g\g_1}g^{\g_1\g_2}g^{\g_2\g_3}\cdots
g^{\g_{l+1}\g}}{(\widetilde{E}_{\g}-E_{\g_1})(\widetilde{E}_{\g}-E_{\g_2})
(\widetilde{E}_{\g}-E_{\g_3})\cdots(\widetilde{E}_{\g}-E_{\g_{l+1}})}
\eeqa Then, we change the starting value of the summation in the
second term of above equation via. the transformation $l\rightarrow
l+1$, we can arrive at \beqa \sum_{\g^\prime \neq
 \g}g^{\g\g^\prime}Q_{\g^\prime\g} &=&\sum_{\g_1\neq \g}\frac{g^{\g\g_1}g^{\g_1\g}}{\widetilde{E}_\g-E_{\g_1}}
 +\sum_{l=2}^\infty\sum_{\g_1,\g_2,\cdots,\g_{l}\neq
\g}\frac{g^{\g\g_1}g^{\g_1\g_2}\cdots
g^{\g_{l}\g}}{(\widetilde{E}_{\g}-E_{\g_1})(\widetilde{E}_{\g}-E_{\g_2})
\cdots(\widetilde{E}_{\g}-E_{\g_{l}})}\nonumber\\
&=& R_\g\left(E_\g-\widetilde{E}_\g\right)=\widetilde{E}_{\g}-E_\g
\eeqa In its last step, Eq. (\ref{ourHEVE}) has been used.
Similarly, using the trick of dummy index transformation, we have
\beqa \sum_{\g^\prime \neq
\g}g^{\g^{\prime\prime}\g^\prime}Q_{\g^\prime\g} &=& \sum_{\g^\prime
\neq
\g}\frac{g^{\g^{\prime\prime}\g^\prime}g^{\g^\prime\g}}{\widetilde{E}_\g
-E_{\g^\prime}}+\sum_{l=1}^\infty\sum_{\g^\prime,\g_1,\g_2,\cdots,\g_l\neq
\g}\frac{g^{\g^{\prime\prime}\g^\prime}g^{\g^\prime\g_1}g^{\g_1\g_2}\cdots
g^{\g_l\g}}{(\widetilde{E}_{\g}-E_{\g^\prime})(\widetilde{E}_{\g}-E_{\g_1})
(\widetilde{E}_{\g}-E_{\g_2})\cdots(\widetilde{E}_{\g}-E_{\g_l})}\nonumber\\
&=&
\left(\widetilde{E}_\g-E_{\g^{\prime\prime}}\right)\left[\sum_{\g^\prime
\neq
\g}\frac{g^{\g^{\prime\prime}\g^\prime}g^{\g^\prime\g}}{\left(\widetilde{E}_\g
-E_{\g^{\prime\prime}}\right)\left(\widetilde{E}_\g-E_{\g^\prime}\right)}\right.\nonumber\\
& &\left.+\sum_{l=1}^\infty\sum_{\g^\prime,\g_1,\g_2,\cdots,\g_l\neq
\g}\frac{g^{\g^{\prime\prime}\g^\prime}g^{\g^\prime\g_1}g^{\g_1\g_2}\cdots
g^{\g_l\g}}{\left(\widetilde{E}_\g-E_{\g^{\prime\prime}}\right)(\widetilde{E}_{\g}-E_{\g^\prime})(\widetilde{E}_{\g}-E_{\g_1})
(\widetilde{E}_{\g}-E_{\g_2})\cdots(\widetilde{E}_{\g}-E_{\g_l})}\right]\nonumber\\
&=&
\left(\widetilde{E}_\g-E_{\g^{\prime\prime}}\right)\left[\sum_{\g_1
\neq
\g}\frac{g^{\g^{\prime\prime}\g_1}g^{\g_1\g}}{\left(\widetilde{E}_\g
-E_{\g^{\prime\prime}}\right)\left(\widetilde{E}_\g-E_{\g_1}\right)}\right.\nonumber\\
& &\left.+\sum_{l=1}^\infty\sum_{\g_1,\g_2,\cdots,\g_{l+1}\neq
\g}\frac{g^{\g^{\prime\prime}\g_1}g^{\g_1\g_2}g^{\g_2\g_3}\cdots
g^{\g_{l+1}\g}}{\left(\widetilde{E}_\g-E_{\g^{\prime\prime}}\right)(\widetilde{E}_{\g}-E_{\g_1})(\widetilde{E}_{\g}-E_{\g_2})
(\widetilde{E}_{\g}-E_{\g_3})\cdots(\widetilde{E}_{\g}-E_{\g_{l+1}})}\right]\nonumber\\
&=&\left(\widetilde{E}_\g-E_{\g^{\prime\prime}}\right)\sum_{l=1}^\infty\sum_{\g_1,\g_2,\cdots,\g_l\neq
\g}\frac{g^{\g^{\prime\prime}\g_1}g^{\g_1\g_2}\cdots
g^{\g_l\g}}{(\widetilde{E}_{\g}-E_{\g^{\prime\prime}})(\widetilde{E}_{\g}
-E_{\g_1})(\widetilde{E}_{\g}-E_{\g_2})\cdots(\widetilde{E}_{\g}-E_{\g_l})}\nonumber\\
&=&\left(\widetilde{E}_\g-E_{\g^{\prime\prime}}\right)Q_{\g^{\prime\prime}\g}-g^{\g^{\prime\prime}\g}
\eeqa Thus Eq.(\ref{secondstep}) has the following form
\beqa
\sum_{\g^{\prime\prime}}\sum_{\g^\prime \neq
 \g}Q_{\g^\prime\g}g^{\g^{\prime\prime}\g^\prime}\ket{\Phi_{\g^{\prime\prime}}}&=&\left(\widetilde{E}_{\g}-E_\g\right)\ket{\Phi_\g}
 +\sum_{\g^{\prime\prime}\neq \g}\left[\left(\widetilde{E}_\g-E_{\g^{\prime\prime}}\right)Q_{\g^{\prime\prime}\g}
 -g^{\g^{\prime\prime}\g}\right]\ket{\Phi_{\g^{\prime\prime}}}\eeqa
Finally, substituting above relation into Eq.(\ref{vstep1}), we
obtain \beq\label{vstepfinal} H\left(\ket{\Phi_\g}+\sum_{\g^\prime
\neq
 \g}
 Q_{\g^\prime\g}\ket{\Phi_{\g^\prime}}\right)=\widetilde{E}_\g\left(\ket{\Phi_\g}+\sum_{\g^\prime
\neq
 \g}
 Q_{\g^\prime\g}\ket{\Phi_{\g^\prime}}\right)\eeq
It immediately goes to our conclusion.

This proof is very easy even for a beginner in quantum mechanics if
he/she is familiar to the method of dummy index transformation. Why
such a complicated problem to solve the stationary Schr\"odinger's
equation in a general quantum system is changed to so simple? It is
not because we can luckily guess this conclusion, but we obtain it
based on our previous works \cite{wangev,wang1}.

In the study on some applications of quantum mechanics, we often
encounter even befall some difficult to understand and calculate
problems, in particular, for open systems because the Hamiltonian is
too complicated to solve its eigenvalue problem. Although the
perturbative solution of SSE has been known, but its general term in
$c$-number form has not been written obviously at our knowledge,
only the Hamiltonian eigenvalues have a general term in the operator
form. As to the Hamiltonian eigenvectors, their time-independent
solution are expressed by Green function or operator in the
potential scatter theory, as well as their dynamical form solution
are written by the time evolution operator or the path integral.
Consequently, we would like to seek a new way even formulism in
order to solve the faced difficulties.

As is understood by us, after giving up the perturbation concept, it
is almost impossible to seek for the exact solution of the SSE
directly. However, we know that the exact solution of dynamical
Schr\"odinger equation can be obtained form the exact solution of
SSE if the Hamiltonian is time-independent. Thus, vice versa, can we
obtain the exact solution of SSE from the exact solution of
dynamical Schr\"dinger equation? While the later in a general system
has been known at least in form via the action of the time evolution
operator. However, its obvious shortcoming is the expression of the
time evolution operator is too formal to be usable directly in the
theory and application. So, the key to seek for the exact solution
of SSE is now changed to how to find out the general and explicit
$c$-number form of time evolution operator. If this aim is arrived
at, this new mentality of our research will have a successful
possibility.

As is well-known, to find the explicit $c$-number function form of
the time evolution operator is a successful linchpin of Feynman path
integral formulism \cite{Feynman}, and to expand the time evolution
operator as a power series is a powerful headstream of Dyson series
(in the interaction picture) \cite{Dyson}. Consequently, our
physical idea \cite{wang1} comes from the combination of Feynman
path integral spirit and Dyson series kernel. We first derive out a
general and explicit $c$-number expression of the time evolution
operator that is different from them as the following: \beq
\label{WangExpression} \ket{\Psi(t)}=\sum_{\g,\g^\prime}A^{\g
\g'}\diracsp{\Phi_{\g^\prime}}{\Psi(0)}\ket{\Phi_\g},\qquad A^{\g
\g'}=\bra{\Phi_\gamma}\e^{-\I H
t}\ket{\Phi_{\gamma^\prime}}=\sum_{l=0}^\infty
A_l^{\gamma\gamma^\prime} \eeq where the final state denoted by
$\ket{\Psi(t)}$, a given arbitrary initial state is denoted by
$\ket{\Psi(0)}$ and the coefficients $A_l^{\gamma\gamma^\prime}$ are
defined by\vskip -0.1in \beqa \label{ADefinition}
A_0^{\gamma\gamma^\prime}= \e^{\I E_{\g}t}\delta_{\g\g'},\qquad
A_l^{\gamma\gamma^\prime}=\sum_{\g_1,\g_2,\ldots,\g_{l+1}}
\sum_{i=1}^{l+1}\frac{\delta_{\g\g_1}\delta_{\g'\g_{l+1}}\e^{-\I
E_{\g_i}t}}{\displaystyle\prod_{j=1,j\neq
i}^{l+1}\left(E_{\g_i}-E_{\g_j}\right)} \prod_{k=1}^l
g^{\g_k\g_{k+1}} \eeqa Note that it is still in a series form
because that there is usually not the compact exact solution (or the
exact solution in the closed form) of SSE for a general quantum
system. It is clear that this expression can be thought of to be
exact in the sense that any approximation means is not used, or that
the whole contributions from all order perturbations are involved if
the perturbation concept is introduced.

It is very interesting to study the relation between our dynamical
exact solution of the dynamical Schr\"odinger equation
(\ref{WangExpression}) and the results from the normal perturbation
theory. By careful calculations and analyses, we found that our
solution (\ref{WangExpression}) can naturally and reasonably
involves partly corrections or contributions from the higher order
perturbations \cite{wang1}, but these corrections are cut out in the
normal perturbation theory, for example, the transition probability
in a very short time as well as the amplitude of nearer states in
the perturbed wave function exist these problems \cite{wang1,wang3}.
Moreover, in the process of the calculations using the solution
(\ref{WangExpression}) we observe two basic facts: (1) The total
Hamiltonian should be split into diagonal and off-diagonal two parts
rather than the perturbed and non-perturbed two parts; (2)
Introducing perturbed parameter should be delayed as possible, had
better do it in the final calculation. So we propose a Hamiltonian
redivision skill and dynamical rearrangement technology in order to
implement our above views. Furthermore, we build an improved scheme
of perturbation theory based on our dynamical exact solution
(\ref{WangExpression}) so that the corrections or contributions from
partly higher order perturbations are involved in the results, and
so our method leads to that the related problems are more accurately
and effectively calculated. Particularly, by analyzing our results,
we propose a conjecture about the rearrangement series of
Hamiltonian eigenvalues and realize a conclusion that the perturbed
series of the Hamiltonian eigenvalues or Hamiltonian eigenvector is
inherently multiple. In other words, the power expansion of
perturbed parameter is not a best method.

As an extension of our frameworks \cite{wang1}, we continue to
present the open system dynamics and the form of Green function and
path integral \cite{wang2}. In the study on them, we further
understand why our results are needed.

Obviously, there are many apparent singular points in the expression
(\ref{ADefinition}) of $A_l^{\gamma\gamma^\prime}$ , but they are
fake in fact. In our manuscript \cite{wang1} this problem has been
fixed by finding their limitations in terms of contraction and
anti-contraction of energy summation indexes. Recently, by
theorizing this method and using partition function \cite{wangev},
we neatly removed all the apparent singular points and arrive at
\begin{equation}\label{PFExpression}
\sum_{\g}\e^{-\I\widetilde{E}{\g}t}=\sum_\gamma A^{\g \g}=\sum_{\g}
\e^{-\I E{\g}t}
    \left\{1+(-\I t)
    \sum_{m=0}^\infty\frac{(-1)^{m}}{(m+1)!}\left[
    \left.\Dif{z}{m}
    \left(\e^{\I z t} R_{\g}^{m+1}(z) \right)
    \right]
    \right|_{z=0}\right\}
\end{equation}
However, more importantly and interestingly, we obtain \beq
\label{PFfinal}\sum_\g \e^{-\I \widetilde{E}_\g t}=\sum_\g
\exp\left\{-\I
\left({E}_\g+\sum_{m=0}^\infty\left.\frac{(-1)^{m}}{(m+1)!}
\left[\Dif{z}{m}R_{\g}^{m+1}(z)\right]\right|_{z=0}\right)t\right\}
\eeq It is proved by expending the partition function in
Eq.(\ref{PFExpression}) into the time power series and verifying the
coefficient power relation. In fact, the form of Eq.(\ref{PFfinal})
has its physics origin rather than the mathematics arbitrariness,
and it is valid in the general quantum systems independent of the
form of Hamiltonian. Therefore, we think that the complete series
expression of Hamiltonian eigenvalues in the general quantum systems
is just:
\begin{equation}\label{EVexpression}
\widetilde{E}_\gamma =
E_\g+\sum_{m=0}^\infty\frac{(-1)^{m}}{(m+1)!}\left[\Dif{z}{m}R_{\g}^{m+1}(z)\right]\Bigg|_{z=0}
\end{equation}
Obviously, it is simply not a summation over the perturbed
parameter, but a series of power of the kernel function $R_\g(z)$ as
well as its derivative at $z=0$. It is completely different from the
normal perturbation theory in its thoughtway. In particular, when a
cut is introduced in a practical calculation, a higher $\lambda^2$
term than the last term of cut part is dropped, but not a higher
$\lambda$ term than the last term of cut part is dropped in the
normal perturbation theory. Moreover, by studying expression of
Hamiltonian eigenvalues (\ref{EVexpression}), we propose a
calculating approach of eigenvalues of arbitrary Hamiltonian via
solving an algebra equation (\ref{ourHEVE}) \cite{wangev}. In
addition, recalling the Richardson's exact solution in the pair
model \cite{Richardson}, we can see that there is a little similar
in form between it and ours because the eigenvalues in the
Richardson's exact solution also obeys an algebra equation, but the
Rechardson's exact solution is only usable in a special system.

Without loss of generality, we can expand the Hamiltonian
eigenvector $\ket{\Psi_\g}$ in the solvable representation spanned
$\{\ket{\Phi_\g}\}$, that is, $\ket{\Psi_\g}=\sum_{\g}
a_{\g\g^\prime}\ket{\Phi_{\g^\prime}}$. For simplicity, we set
$a_{\g\g}=1$ for a given $\g$ since that $\ket{\Psi_\g}$ only can be
determined to the difference from a normalization factor. Then,
acting $\bra{\Phi_\g}$ on Eq.(\ref{HEE}), we have
$\widetilde{E}_\g=E_\g+\sum_{\g^\prime} g^{\g\g^\prime}
a_{\g\g^\prime}$. It is easy to obtain its solution $
a_{\g\g^\prime}=\delta_{\g\g^\prime}+Q_{\g^\prime\g}\left(1-\delta_{\g\g^\prime}\right)$
if the total Hamiltonian eigenvalues is a solution of
Eq.(\ref{ourHEVE}). This derivation also verifies the result that
the total Hamiltonian eigenvalues obey such an algebra equation
(\ref{ourHEVE}).

Therefore, our solution of the stationary Schr\"odinger's equation
in the general quantum systems is given originally based on our
formal framework on quantum mechanics in the general quantum system,
in particular, the conclusions in our paper \cite{wangev}, but it
does not come from a guess although we can drop the details of our
derivation. It can be thought of to be exact in the sense that any
approximation means is not used, or that whole corrections and
contributions from all order perturbations are involved if the
perturbation concept is introduced. Breaking the accustomed
mentality in the normal perturbation theory, finding the right
division principle of Hamiltonian, using the more explicit $c$
number expansion form of time evolution operator, dealing skillfully
with and reasonably rearranging the perturbed series are advantages
of our formulism \cite{wang1}. They bring us successfully to arrive
at our purpose, and are also the reasons why our exact series
solution of SSE can have more and new physics content.

For convenience in the theoretical derivation and efficiency in the
practical calculation, let us rewrite our solution in a more compact
form. Firstly, introducing a project operator \beq
P_\gamma=I-\ket{\Phi_\g}\bra{\Phi_\g}\eeq Here, $I$ is an identity
matrix and defining a limited operator of $H_1$ \beq
\overline{H}_{1\g}=P_\g H_1P_\g\eeq we have \beq
\bra{\Phi_{\g_1}}\overline{H}_{1\g}\ket{\Phi_{\g_2}}=\bar{g}_\g^{\g_1\g_2}=\left\{\begin{array}{ll}g^{\g_1\g_2}&
({\rm if} \quad \g_1,\g_2\neq\g)\\
0 & ({\rm otherwise})\end{array}\right.\eeq where we have used the
facts that $P_\g\ket{\Phi_\g}=0$ and $\bra{\Phi_\g}P_\g=0$. It
implies that the elements of the $\g$-th row and $\g$-th column of
$\overline{H}_{1\g}$ matrix are zero, the other elements are same as
the $H_1$ matrix in the $H_0$ representation. Then, we again define
a revised Green operator \beq
\widetilde{G}_\g=\frac{1}{\widetilde{E}_\g-H_0}\eeq we arrive at
\beqa
Q_{\g^\prime\g}&=&\frac{1}{\widetilde{E}_\g-E_{\g^\prime}}g^{\g^\prime\g}
+\sum_{l=1}^\infty\sum_{\g_1,\g_2,\cdots,\g_l}\frac{1}{\widetilde{E}_{\g}-E_{\g^\prime}}g^{\g^\prime\g_1}\frac{1}{\widetilde{E}_{\g}
-E_{\g_1}}\bar{g}^{\g_1\g_2}\frac{1}{\widetilde{E}_{\g}-E_{\g_2}}
\nonumber \\
&
&\cdots\frac{1}{\widetilde{E}_{\g}-E_{\g_{l-1}}}\bar{g}^{\g_{l-1}\g_l}\frac{1}{\widetilde{E}_{\g}-E_{\g_l}}
g^{\g_l\g}\nonumber\\
&=&\bra{\Phi_{\g^\prime}}\widetilde{G}_\g H_1\ket{\Phi_\g}+
\bra{\Phi_{\g^\prime}}\widetilde{G}_\g
H_1\sum_{l=0}^\infty\left(\widetilde{G}_\g\overline{H}_{1\g}\right)^{l}\widetilde{G}_\g
H_1\ket{\Phi_\g}\eeqa Note that $H_1$ matrix is off-diagonal, that
is, $g^{\g\g}=0$ for any $\g$. If we focus on such quantum systems
that the following identity is valid \beq\label{condition}
\frac{1}{I-\widetilde{G}_\g
\overline{H}_{1\g}}=\sum_{l=0}^\infty\left(\widetilde{G}_\g\overline{H}_{1\g}\right)^{l}\eeq
we obtain \beq
Q_{\g^\prime\g}=\bra{\Phi_{\g^\prime}}\widetilde{G}_\g
H_1\ket{\Phi_\g}+ \bra{\Phi_{\g^\prime}}\widetilde{G}_\g
H_1\frac{1}{I-\widetilde{G}_\g \overline{H}_{1\g}}\widetilde{G}_\g
H_1\ket{\Phi_\g}\eeq In fact, the same skill has been used
extensively in the normal Green operator method, so it is still
general enough. Therefore, the eigenvector of our exact series
solution is rewritten as an operator form \beqa \label{evectorof}
\ket{\Psi_\g}&=&\ket{\Phi_\g}+\sum_{\g^\prime\neq
\g}\ket{\Phi_{\g^\prime}}\bra{\Phi_{\g^\prime}}\widetilde{G}_\g
H_1\left[I+\frac{1}{I-\widetilde{G}_\g
\overline{H}_{1\g}}\widetilde{G}_\g
H_1\right]\ket{\Phi_\g}\nonumber\\
&=&\ket{\Phi_\g}+P_\g\widetilde{G}_\g
H_1\left[I+\frac{1}{I-\widetilde{G}_\g
\overline{H}_{1\g}}\widetilde{G}_\g H_1\right]\ket{\Phi_\g}\eeqa
Comparing with the Green operator method \beq
\ket{\Psi_\g}=\ket{\Phi_\g}+G_\g H_1\ket{\Phi_\g} \eeq it
immediately follows that the complete Green operator can be
expressed by \beq G_\g=P_\g\widetilde{G}_\g +P_\g\widetilde{G}_\g
H_1\frac{1}{I-\widetilde{G}_\g \overline{H}_{1\g}}\widetilde{G}_\g
\eeq When one tries to use it to the potential scatter problems, the
boundary condition has to be considered. The related work is put in
our another manuscript (in prepare).

Similarly, the equation satisfied by the eigenvalues of our exact
series solution can be rewritten as \beq\label{evalueof}
\bra{\Phi_\g}H_1\frac{1}{I-\widetilde{G}_\g
\overline{H}_{1\g}}\widetilde{G}_\g
H_1\ket{\Phi_\g}=\bra{\Phi_\g}H_1\widetilde{G}_\g\frac{1}{I-
\overline{H}_{1\g}\widetilde{G}_\g}
H_1\ket{\Phi_\g}=\widetilde{E}_\g-E_\g\eeq This expression will
largely advance the efficiency in the numerical calculations. Some
examples studied by us are in progressing.

It must be emphasized that our solution is a universal, systematical
and programmable solution that is independent of the form of
Hamiltonian. The preconditions of our solution is that the
Hamiltonian contains a solvable part, which is the same as the
perturbation theory and the Green function method. However, we need
that the Hamiltonian is spilt into a diagonal $H_0$ and a
off-diagonal $H_1$, and it is always made. In fact, our solution can
be thought of an Ansatz. Direct verification indicates its validity,
and so its preconditions are looser. Of course, we have used the
assumption of existence and uniqueness of solution of SSE for the
general quantum systems.

It is easy to verify that our results is consistent with ones in the
normal perturbation theory if one expands our expressions of
Hamiltonian eigenvectors and eigenvalues according to the order of
perturbed parameter for the known lower order forms. It implies not
only that our solution contains the perturbation solution, but also
our solution have more physics content. As an exact series solution,
our solution has a neat form of general term and involves the
contributions from all order perturbation if the perturbation
concept is needed. Actually, our solution is inherently a
non-perturbation solution. Comparing with the the Green function
method, our solution does not need to solve the differential
equation, and comparing with the directly diagonalized method, our
expression is in a analytical form rather than a numerical form.
Therefore, our solution is more suitable to the theoretical
derivation and proof. Moreover, the convergence of our solution is
considered in terms of the algebra equation satisfied by the
Hamiltonian eigenvalues. In other words, the existence of solution
of this equation implies that the series of our solution can
converge to a finite number.

A universal form has to pay the cost. A universal solution is
unnecessary optimal for a special system. From the universality to
speciality it still needs to study. In a concrete system, the form
of our exact solution should be and can be simplified. In
particular, we can not claim that our solution has the optimal
efficiency in the numerical calculation for some special systems.
This situation ought to be understandable, for example, the direct
diagonalization method to find the eigenvectors and eigenvalues of
Hamiltonian in some simple systems has the higher efficiency in the
numerical calculations than the perturbed solution method to do
this. However, ones cannot use this evidence to deny the importance
of perturbative solution as well as perturbation theory. In fact,
the direct diagonalization method is a good method limited within
the numerical calculations. If there were not the perturbation
theory and perturbative solution, will quantum mechanics have today
achievement? The academic value and scientific significance of our
exact series solution should be judged based on whether our solution
contains more physics content than the perturbative solution.

Apparently, our solution seems a rearranging summation when the
perturbation concept is introduced, but essentially it reveals the
physics nature in a quantum system. This is an important reason why
a complicated problem to solve the stationary Schr\"odinger equation
in a general quantum system is changed to so simple. In fact, our
solution directly comes from quantum dynamics and it has the
completeness, orderliness and clearness. In practical calculations,
they also provide a physics method how to choose part corrections or
contributions from higher order perturbations, which is able to
simplify the calculation and lead the result more precise.
Therefore, our solution is not trivially equivalent to the normal
perturbative solution when the perturbation concept is introduced
because that the reasonably dynamical rearrangement feature of our
solution leads to the effects from the different cut approximation.
Two examples have been presented in Refs.\cite{wangev,wang3}. In
special, using the expressions of our exact solution of SSE
(\ref{evectorof}) and (\ref{evalueof}) for many interesting systems,
which satisfies the condition (\ref{condition}), we can obtain some
encouraging results.

By comparison and analyses, our results are helpful for theoretical
derivations and practical calculations since its universal,
analytical and exact features. However, it must be pointed out that
we expect that our exact series solution of SSE has more important
and interesting conclusions and applications. It will be more
clearly seen the academic value and scientific significance of our
exact series solution of SSE.

For simplicity, we only consider the non-degeneracy and discrete
case here, but our derivation can be extended to the degeneracy
and/or continuous case. At present, we are studying some concrete
examples in order to further verify the validity of our results and
reveal the ability of our approach. We are sure that it will bring
many interesting conclusions when our frameworks
\cite{wangev,wang1,wang2,wang3} are extended to the time-dependent
cases and quantum field theory.

\medskip

I am very grateful for Zhou Li for her irreplaceable contributions
from the expression and calculation of Hamiltonian eigenvalues in
the general quantum systems \cite{wangev}. This work was supported
by the National Natural Science Foundation of China under Grant No.
10975125.


\end{document}